# The Layer 0 Inner Silicon Detector of the D0 Experiment

R. Angstadt[1], L. Bagby[1], A. Bean[8], T. Bolton[7], D. Buchholz[2], D. Butler[1], L. Christofek[8], W.E.Cooper[1], C.H. Daly[3], M. Demarteau[1], J. Foglesong[1], C.E. Gerber[9], H. Gonzalez[1], J.Green[1], H. Guldenman[3], K. Hanagaki[1], K. Herner[5], J. Howell[1], M. Hrycyk[1], M. Johnson[1], M. Kirby[2], K. Krempetz[1], W.Kuykendall[3], F.Lehner[4], R. Lipton[1], H.J.Lubatti[3], D. Markley[1], M. Matulik[1], R.L. McCarthy[5], A. Nomerotski[1], D. Olis[1], Y. Orlov[1], G.J. Otero y Garzón[9], M. Roman[1], R. Rucinski[1], K. Schultz[1], E. Shabalina[9], R.P. Smith[1,10], D. Strom[2], R.D. Taylor[7], D. Tsybychev[5], M. Tuttle[3], M. Utes[1], J. Wang[3], M. Weber[1], T. Wesson[1], S.W.Youn[2], T.Zhou[3], A. Zieminski[6,10]

1 Fermi National Accelerator Laboratory, Batavia, Illinois 60510, USA
2 Northwestern University, Evanston, Illinois, 60208, USA
3 University of Washington, Seattle, Washington, 98195, USA
4 Physik Institut der Universität Zürich, Zürich, Switzerland
5 State University of New York, Stony Brook, New York, 11794, USA
6 Indiana University, Bloomington, Indiana, 47405, USA
7 Kansas State University, Manhattan, Kansas, 66506, USA
8 University of Kansas, Lawrence, Kansas, 66045, USA
9 University of Illinois at Chicago, Chicago, Illinois, 60607, USA
10 Deceased

**Corresponding author.**

Henry J. Lubatti, Professor
Department of Physics
University of Washington, Box 361560
Seattle, WA 98195-1560
U.S.A.

Tel: 1 206 543 8964
Fax: 1 206 685 9242

E-mail: lubatti@u.washington.edu

## Abstract

This paper describes the design, fabrication, installation and performance of the new inner layer called Layer 0 (L0) that was inserted in the existing Run IIa Silicon Micro-Strip Tracker (SMT) of the D0 experiment at the Fermilab Tevatron p̄p collider. L0 provides tracking information from two layers of sensors, which are mounted with center lines at a radial distance of 16.1 mm and 17.6 mm respectively from the beam axis. The sensors and readout electronics are mounted on a specially designed and fabricated carbon fiber structure that includes cooling for sensor and readout electronics. The structure has a thin polyimide circuit bonded to it so that the circuit couples electrically to the carbon fiber allowing the support structure to be used both for detector grounding and a low impedance connection between the remotely mounted hybrids and the sensors.

## 1. Introduction

The design and installation of a new inner layer for the existing silicon microstrip tracker (SMT) was driven by a desire to improve impact parameter resolution, increase b-tagging efficiency and, in general, provide more robust tracking and pattern recognition at higher luminosities. Achieving these goals required a layer of sensors at small radius while at the same time keeping the number of radiation lengths to a minimum and conforming to the existing Run II detector geometrical constraints. L0 had to be installed through the existing SMT aperture of 22.90 mm radius. This limited the envelope of the detector, including signal cables, etc. to 22.02 mm radius. The desired goals and constraints led us to choose carbon-fiber for the support structure as a way to achieve a very stiff, low-Z, support structure that could be relatively easily made into the desired geometry and within required tolerances.

L0 contains 48 radiation-hard, single-sided silicon sensors and 48 hybrid electronic circuits, each containing two SVX4 [2,3] chips all of which are mounted on a high elastic modulus, low-mass carbon-fiber composite support structure that contains PEEK (polyetheretherketone) cooling tubes. Figure 1 shows the completed L0 structure prior to installation into the SMT. Because of its small diameter and long length, the detector required both innovative mechanical and electrical designs.

The support structure is constructed using carbon-fiber/epoxy composite (CF) tubes made from high elastic modulus carbon fiber. Four castellated hexagonal outer tubes on which sensors and hybrids are mounted are glued to one dodecahedral inner tube to achieve the required mechanical shape and stiffness.  This geometry maximizes the coverage azimuthally and minimizes the radial distance from the nominal interaction point. In the hybrid region the outer surfaces are simple hexagonal tubes. The support structure design is shown in Figure 2. The entire support structure is 1660 mm long, centered at z=0[1] (the beam crossing center), with sensors covering a z-range of ± 380 mm. Eight 300 µm thick $p^+$ on n single-sided sensors (Hamamatsu Photonics, Inc.) are mounted on each of the six faces of the support tube with three rows at 16.1 mm and three at 17.6 mm from the beam axis.

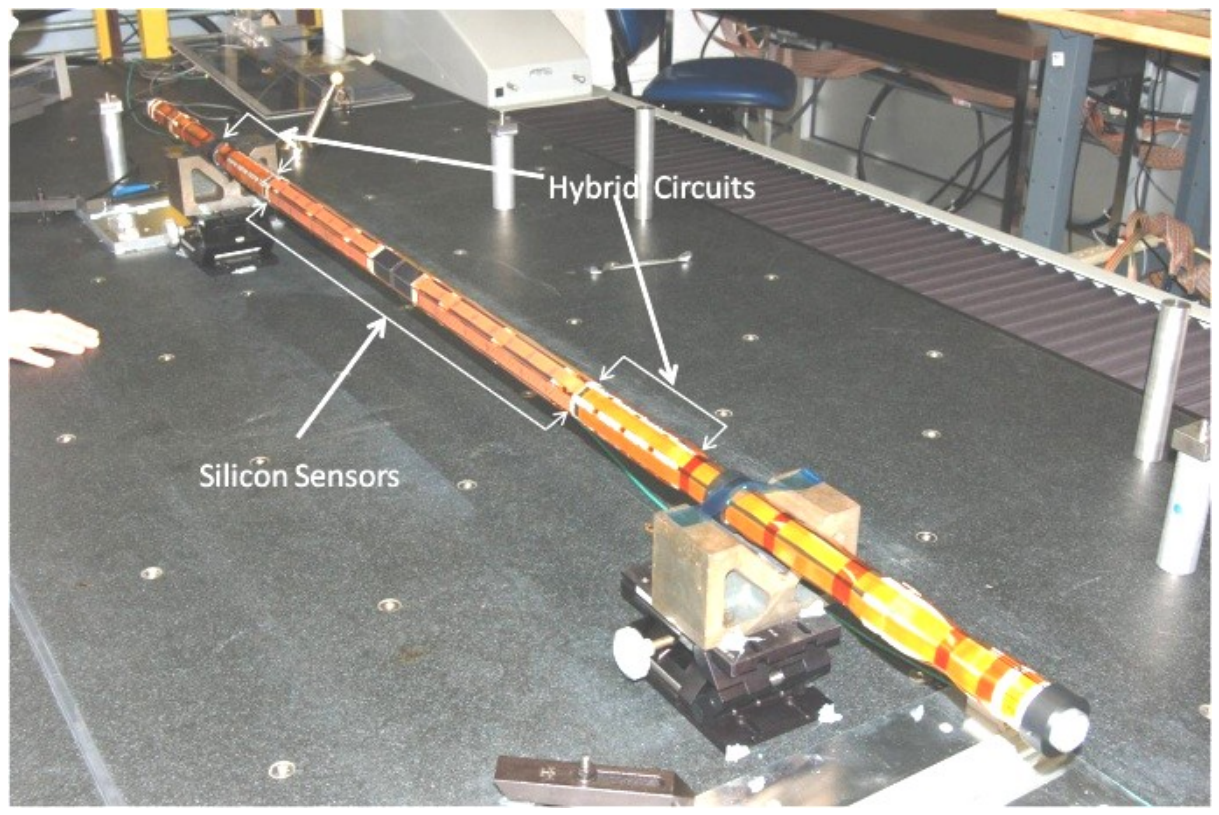

Figure 1. Completed Layer 0 detector.

To limit material in the detector volume, hybrid circuit boards for readout were mounted at each end of L0 and connected to the sensors with fine pitched copper-on-kapton cables.  The capacitance of these cables adds to the sensor capacitance which results in a signal to noise ratio of about 15 to 1.  Thus, the detector had to be designed to minimize common mode noise.

.

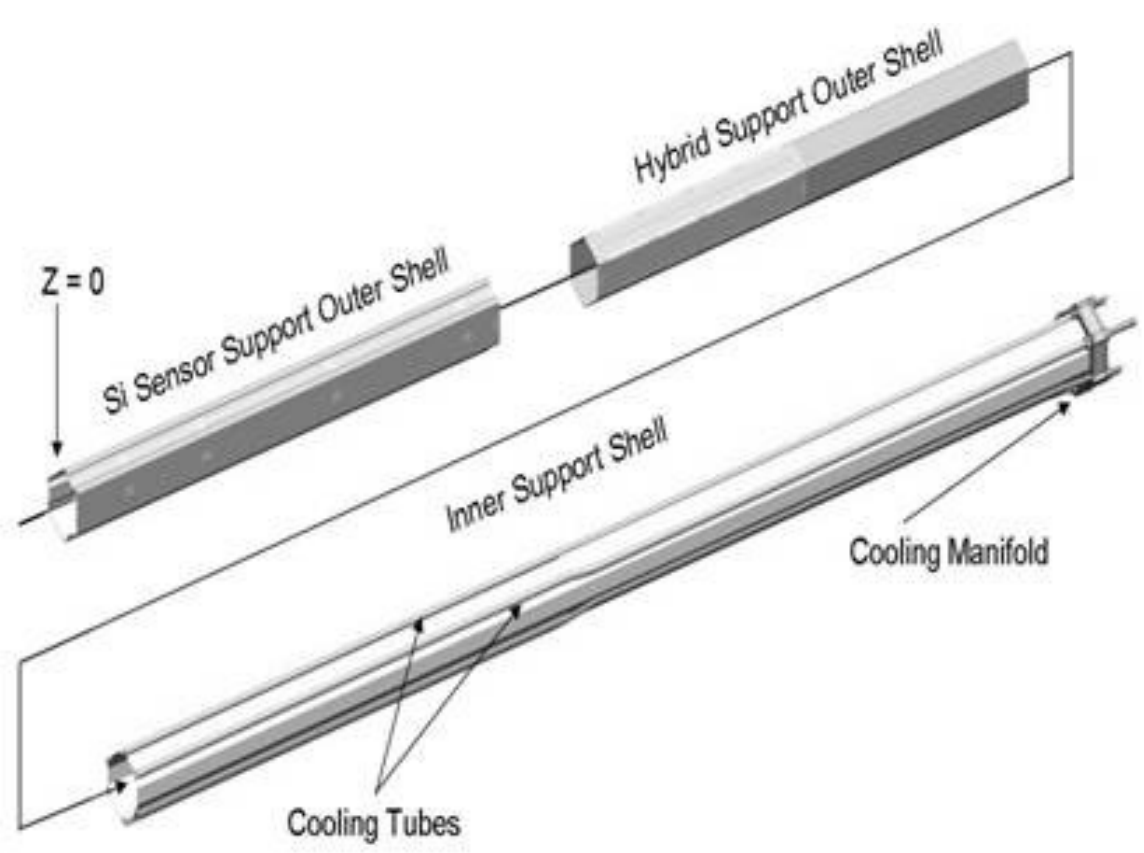

Figure 2. Exploded view of the L0 shell assembly (half structure shown).

[1] The z-direction is the longitudinal axis of the detector, parallel to the proton line of flight.

This was done by exploiting the electrical conductivity of carbon fiber. A thin polyimide printed circuit covered with a mesh ground plane was co-cured with the carbon fiber structure so that there was good electrical contact between the ground mesh and the carbon fibers. This made the support structure part of the grounding system for the detector and it also provided a low impedance connection between the sensor and the hybrid circuits at high frequencies

The following sections present a detailed description of the mechanical design and the methods used for construction. Section 2 describes the mechanical design, analysis and testing. Section 3 describes construction and Section 4 the sensors and testing. Sections 5 and 6 describe the overall electrical design including the grounding and shielding as well as descriptions of the individual components. Section 7 describes installation of L0 in D0 and Section 8 discusses the performance of L0.

## 2. Mechanical Design and Construction

### 2.1 Mechanical Design Constraints

The L0 silicon detector is an addition to the existing D0 Run IIa silicon tracker and had to fit in the space constraints defined by the existing silicon tracker and the new beryllium beam pipe that had been previously procured[2]. The beam pipe has an outer diameter of 29.5 mm in the center beryllium region and an outer diameter of 30.5 mm at the two outside stainless steel flanges. Membranes in the support structure of the original silicon tracker have an aperture of 45.8 mm in diameter. The L0 detector needed to fit within this small annular space with adequate clearance for installation. The length from outer membrane to outer membrane of the existing silicon tracker is 1660 mm.

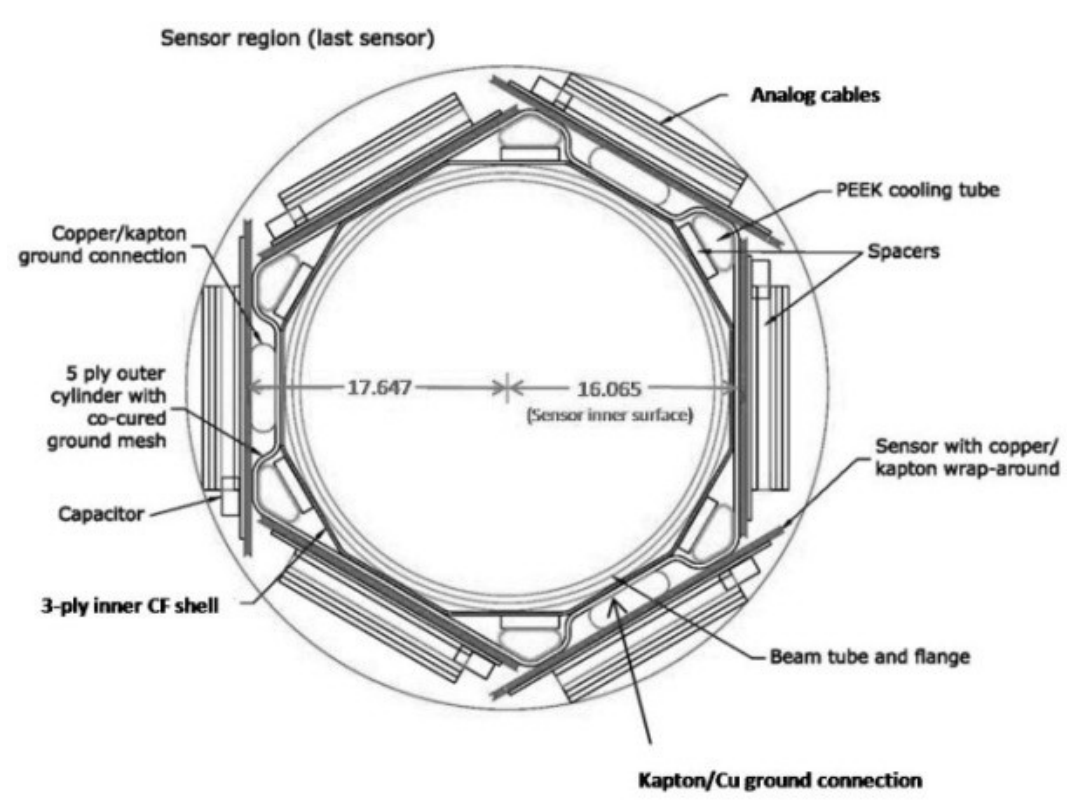

Figure 3. Cross section of detector assembly.

[2] The beam pipe was originally procured as part of the Run IIb upgrade project [2] which included a replacement for the entire SMT. The full SMT upgrade was subsequently canceled and replaced by Layer 0. The L0 detector described herein benefited from the R and D for the original Run IIb SMT upgrade.

The silicon sensors need to be located as physically close to the collision point as possible and be arranged to provide nearly a 360° azimuthal coverage about the beam pipe. A transverse cross section of the detector assembly is shown in Figure 3.
As indicated, the sensors are arranged in an overlapping pattern around the circumference of the beam pipe, so as to intercept as many particles emanating from the collision point as possible.

The read-out chips could not be mounted on the sensors because of the small existing SMT radial aperture, a need to limit material within the tracking volume and to use an existing hybrid design. They were mounted on the hybrid support shells (figure 2) at the ends of the sensor region. This required specially designed polyimide flex cables to carry the analog signals to the read-out chips. The choice of sensor length was driven by approximately equalizing occupancy and maximizing the signal to noise ratio. Matching the shorter sensors with the longer analog flex cables approximately equalized the sensor and cable capacitance, yielding an optimum signal to noise ratio.

The detector support structure had to be designed to minimize radiation length. This constraint implied that the structure should be composed of materials with as low an atomic number as possible, and furthermore, the walls of the structure needed to be as thin as possible. Accurate positioning of the sensors was critical and therefore a very stiff support structure was needed to minimize deflections due to body forces or other external loads experienced during operation. DC electrical isolation between the sensor bias voltage system and the support structure also had to be maintained. Finally, the silicon sensors should be operated at a temperature of less than -5°C to improve the signal to noise ratio and to maximize useful detector life. When energized, the sensors and associated circuitry dissipate about 18 W. Therefore active cooling of the structure is required to maintain -5°C during operation.

## 2.2 Support Structure Material Selection

The need to minimize radiation length, the high stiffness requirement and fabrication issues dictated the use of a high performance carbon fiber/epoxy composite. K13C2U fibers[3] were selected for three reasons. First, the use of ultra-high modulus fibers resulted in a composite material system with very high stiffness. Second, composites produced using these fibers possess very high thermal conductivities relative to composites produced using lower modulus fibers. This was important because preliminary numerical analyses had indicated that a high thermal conductivity would allow us to achieve the desired silicon sensor operating temperature of -5°C. Third, high modulus carbon fiber has high electrical conductivity and this ensures good, robust grounding provided good contact can be made to the sensors and readout electronics.

[3] Mitsubishi Chemical

## 2.3 Design and Analysis of the Support Structure

The dimensions and shape of the shells that made up the support structure were determined by a detailed iterative stress/strain analysis using the ANSYS® finite element analysis (FEA) system. The number and orientation of the layers of carbon fiber/epoxy were optimized to produce a structure that was light and very stiff in bending. The structure also needed to have sufficient circumferential stiffness to minimize any out of round deflections both in use and, especially, during installation of the silicon sensors. Significant local force is applied to squeeze out the epoxy used to mount the sensors and to hold the sensor precisely in the desired position as this epoxy cured. The analysis showed that we needed a temporary internal support inside the shells during this process.

These analyses were based on effective laminate properties, which were themselves predicted using unidirectional ply properties and classical lamination theory[4] (CLT). A limited experimental study was performed to measure ply properties and also to verify that CLT could be used to predict multi-angle laminate properties to within engineering accuracies. A detailed description of the FEA work can be found in [4]

## 2.4 Stacking Sequences and Material Properties

Different stacking sequences were used in the outer castellated and hexagonal tubes and the dodecagonal inner tube. Fiber angles were referenced to the longitudinal axis, e.g., fibers within a 0° ply were parallel to the long axis (z axis) of the tubes. Symmetric stacking sequences were used throughout the structure to eliminate thermal warping. As is evident from Figures 2 and 3, fibers in the outer castellated tubes were required to conform to several small-radius corners. Ply fiber angles were limited to ±20° to avoid fiber fracture at and near these corners. A stacking sequence of +20/-20/0/-20/20° was selected for use in the outer castellated tube. The outer tube therefore possessed high axial and shear stiffness, but very low circumferential stiffness. Since the inner hexagonal tube was produced without sharp corners, fiber angles of any value could be used in this tube. A stacking sequence of 0/90/0° was selected, which enhances the circumferential stiffness once the cylinders are combined with each other while still providing excellent longitudinal stiffness.

## 2.5 Mechanical Design of the Support Structure

The L0 support structure is divided into three major longitudinal regions (Figure 2). The central region, occupying the central 760 mm of the structure, has an outer shell contour which provides precision mounting surfaces for the silicon sensors. The other two regions are the 450 mm long hexagonal hybrid mounting outer shells at each end of the structure. The coolant distribution manifold assemblies at each end also act as the support and connecting points for L0. All of these components are connected together via a long dodecagonal inner shell. The inner shell spans the full length of the support structure (1660 mm) and has an inscribed diameter of 30.7 mm. An outer layer of polyimide with copper mesh and gold plated contacts is co-cured with the outer shells giving a total outer shell cured thickness of 325 μm. The sensor mounting outer shell has a six sided castellated shape that allows phi overlap of sensors to be provided by mounting sensors at two different radial locations. Preformed polyimide/Cu jumpers, 25μm thick, oblong shape and 10 mm z-extent, are used at each larger radius sensor location to connect the gold plated contacts of the sensor shell to the grounding contacts on the underside of the sensor assemblies. In the design of the shape of the inner and outer shells, spaces were reserved for cooling tubes. Figure 3 shows a cross section view of the support structure in the sensor region. The cooling tubes are an integrated part of the support structure. The coolant (a mixture of approximately 70% water and 30% ethylene glycol) flows through the cooling system at a gauge pressure of -0.21 bar to prevent leakage. Special thin wall (100 μm) triangular shaped cooling tubes, extruded from PEEK, which fit the space between the hexagonal outer shells and the inner shell were obtained from TexLoc Inc[5]. It is important to support the cooling tubes on all three flat surfaces to prevent them from collapsing due to the sub-atmospheric coolant pressure. In the sensor region, a spacer made of Rohacell® IG 51 closed-cell, rigid foam is used to bring the cooling tube in contact with the outer shell corners. In the hybrid region, the cooling tubes are directly bonded to the inner and outer shell. The six cooling tubes run the entire length of the structure and connect into the cooling distribution manifolds at each end. The coolant flows in the same direction through all six cooling tubes. Figure 4 shows the complete cooling manifold and an exploded view of the Unigraphics® CAD model of the cooling manifold. Because the cooling tubes need to spread radially in order to connect into the manifold, a small ramp made of Rohacell foam is used to space them out and a cover is bonded to the cooling tubes to protect them. Due to the intricate shape and size of the distribution manifold, it is made of five parts that are bonded together. The parts, all made from PEEK, are bonded together as shown in Figure 4. Three cooling nozzles at each end connect to the coolant supply and return lines.

[4] Classical lamination theory (CLT) is the basic design tool for evaluating different laminates when experimental data are not available. CLT can be used to combine properties and the orientation of each ply in a predetermined stacking sequence to predict the overall performance characteristics for a laminate.

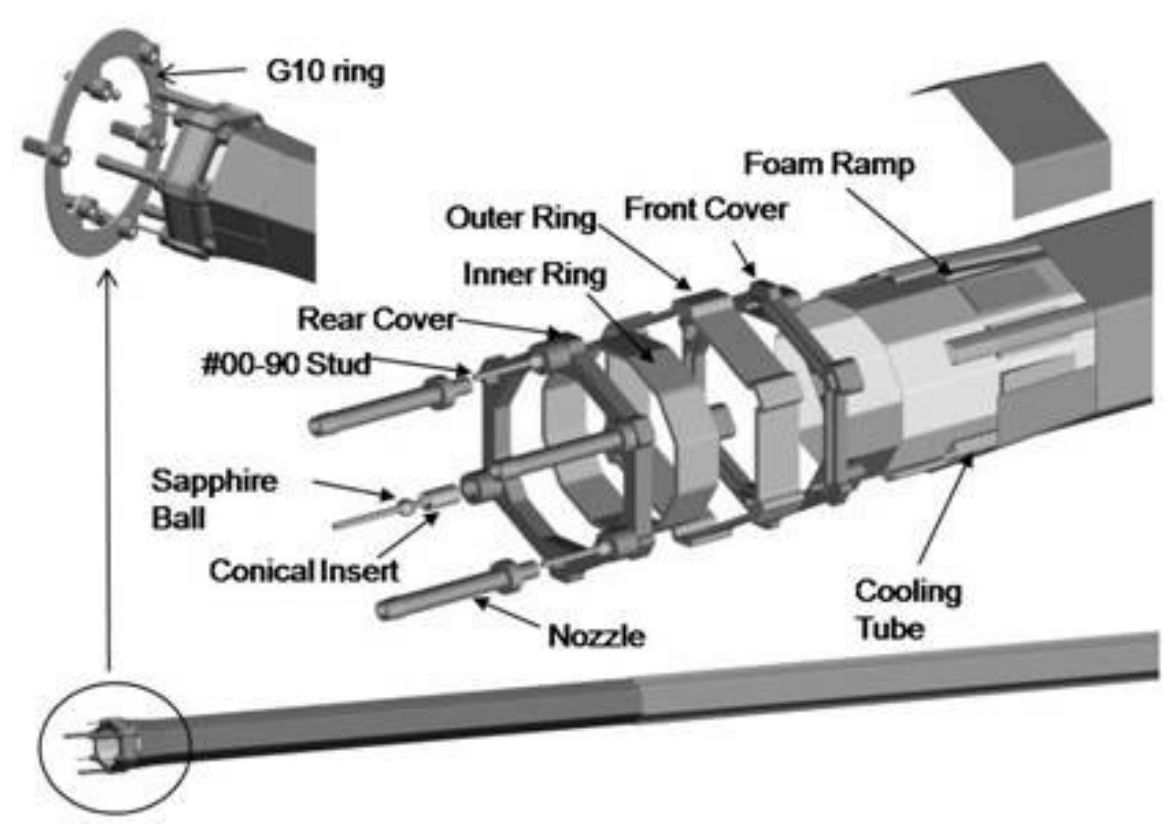

Figure 4. Cooling manifold and mounting system.

---

[5] Parker-Texloc, 4700 Lone Star Blvd,. Fort Worth, Texas 76106, USA

All sensor and hybrid mounting surfaces of the assembled L0 support structure were measured on a coordinate measuring machine (CMM) with a touch probe to establish a structure-based coordinate system for mounting the sensors and hybrids. Measurements of deflection of the structure as a function of a centrally applied load tested structural integrity and allowed compensation for the effects of component weights during installation of the sensors, analog cables and hybrid modules. All sensors and hybrids were installed with their mounting plane oriented horizontally. A rotary encoder measured azimuthal orientation of the structure at each step. Each sensor at a given azimuth was aligned with the aid of a video probe so that, once all parts had been mounted, they would be correctly positioned both transversely and longitudinally. Repeatability of these measurements was approximately ±2 μm transversely and ±7 μm longitudinally. Cylindrical survey monuments with optical fiducials allowed touch probe and video probe measurements to be correlated.

Additional reference spheres, which were temporarily glued to the structure, allowed the coordinate system to be re-established in each azimuthal orientation. After all components had been installed, final sets of optical probe measurements were made with multiple azimuthal orientations of the structure to establish as-built sensor locations and gravitational deflections of the assembly. Those measurements included a determination of the positions of the L0 mounting spheres at each end of the L0 structure.

The mounting connections at each end of L0 are done using a pair of intermediate rings made from G10 that incorporate precision conical sockets. Each of these sockets mates with an identical socket on the end face of each cooling manifold. A precision sapphire sphere fits between the two conical sockets. This system provides a precise, repeatable connection at three points on each end of L0. The relation of the G10 ring and the manifold is shown in Figure 4. There are three other similar cone/sphere locations on the G10 ring. These are used to make the support connection to the existing membranes at each end of the SMT. This requires very precise location of the conical inserts in all of the mating parts and was accomplished by making a master part that was CNC machined from cast aluminum tooling plate from which all production parts were cloned. The conical inserts fit into oversize holes and are glued in place while the new part and the master part are held together using the sapphire spheres to dictate the precise location of the inserts. The relative location of the inserts in the manifolds at each end of L0 was set using a special, precision assembly jig. Conical inserts in the angle brackets at each end of the jig were precisely located by cloning from the master part.

## 2.6 Mechanical Analysis and Testing

Figure 5 shows a comparison of the FEA calculated deflection versus the measured deflection for the completed device under a 100 gram test load. The agreement is good. Calculated stresses in the structures are very low and are not a significant issue. The one circumferential layer in the 3 layer inner shell provides sufficient stiffness against out of round deflection and it also results in a very low circumferential thermal expansion coefficient. The thermal FEA results for L0 show that the sensor temperatures are maintained at well below -5°C and that a very low flow rate of the coolant is required to maintain the desired temperature. The separately mounted hybrid chips have a much less stringent temperature requirement which is easily met.

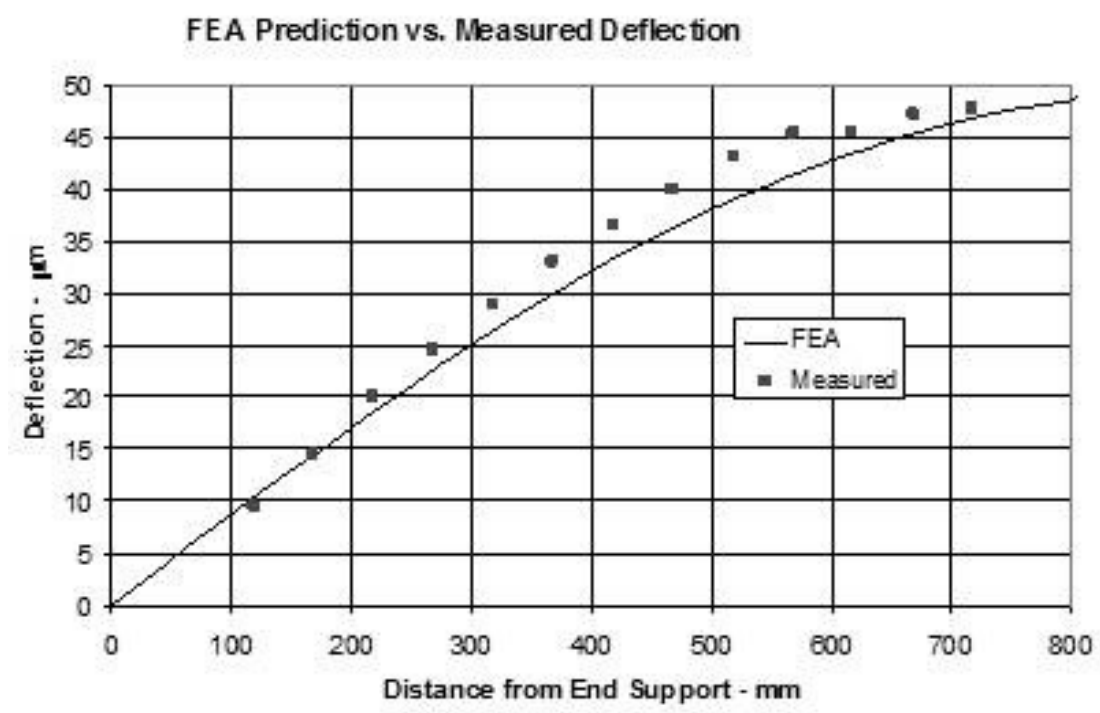

Figure 5. Comparison of the FEA prediction with the measured deflection under a single, central 100 gram load.

## 3. Support Structure Fabrication

The fabrication of various composite components, and their subsequent assembly into the final support structure, is presented in this section. Components include inner and outer shells, electrical grounding circuits, cooling tubes and manifolds, and precision ball mounts. Topics covered include tooling, hand lay-up, vacuum/autoclave curing, CNC machining and epoxy bonding of the assembly.

Fabrication of the carbon fiber support structure required making three separate carbon fiber structures: a full length inner shell and two pairs of outer shells that support the silicon sensors and hybrids (see figure 2). The shells were manufactured by wrapping carbon fiber pre-preg material on steel mandrels and curing in an autoclave. The required dimensional tolerances were achieved by a combination of very precise mandrels, control of parameters during the wrapping of the pre-preg material and precise assembly tooling used in epoxy bonding the inner and outer shells to form a single structural unit.

In all, three mandrels were fabricated. The two 450 mm mandrels for the sensor and hybrid regions were made from steel using standard CNC machining techniques. The 1700 mm mandrel for the inner shell required special techniques. Because of its length and required tolerance control we used NAK-55[6] steel, which is processed to have minimal internal stresses that can result in warping as material is removed. During machining the inner shell mandrel was supported at the two ends and at 6 adjustable intermediate points. The adjustable points made it possible to compensate after each cut for any dimensional changes resulting from internal stress relief. This resulted in a measured maximum total run-out of 18µm and a ±7µm deviation from the nominal distance across the flats (30.68mm).

[6] International Mold Steel, Inc.,Florence, KY 41042

### 3.1 Material Lay-up

3.1.1 Outer Shells

The castellated outer shells on which the sensors are mounted presented the greatest challenge. The mounting surfaces needed to be smooth, flat and within the designed radial envelop (see Figure 3). This required control of material as it was wrapped around the mandrel to ensure it adhered to the surface features of the mandrel with no wrinkling, overlap of ply seams or bridging of mandrel features. The latter is critical to ensuring the desired geometry and performance of the structure. We found that by pre-laminating the ±20º layers on a flat surface we could better control fiber orientation and prevent fibers from separating from the backing tape on the pre-preg during wrapping. This made it possible to achieve a wrinkle-free wrap leading to more reproducible results. The ±20º angles were defined by cutting the sheet of material at this angle so that the wrapping was a straight forward circumferential wrap.

During the wrapping, full length clamping bars that attach to collars were used to secure each ply against each flat surface on the mandrel (see figure 6). Silicon pads with a profile matching the mandrel surface are attached to the clamping bars resulting in a compressible surface that evenly distributes the load. The carbon fiber material is initially aligned with the mandrel axis. Then, a clamping bar with release fabric on the contact surface is attached and this is repeated as one goes around the cylinder. Pulling the release fabric taut after securing the clamps provides a smoothing force that works the material around the corners of the mandrel. The next layer of carbon fiber is done in the same way with each clamping bar removed and replaced as the material is worked around the circumference.

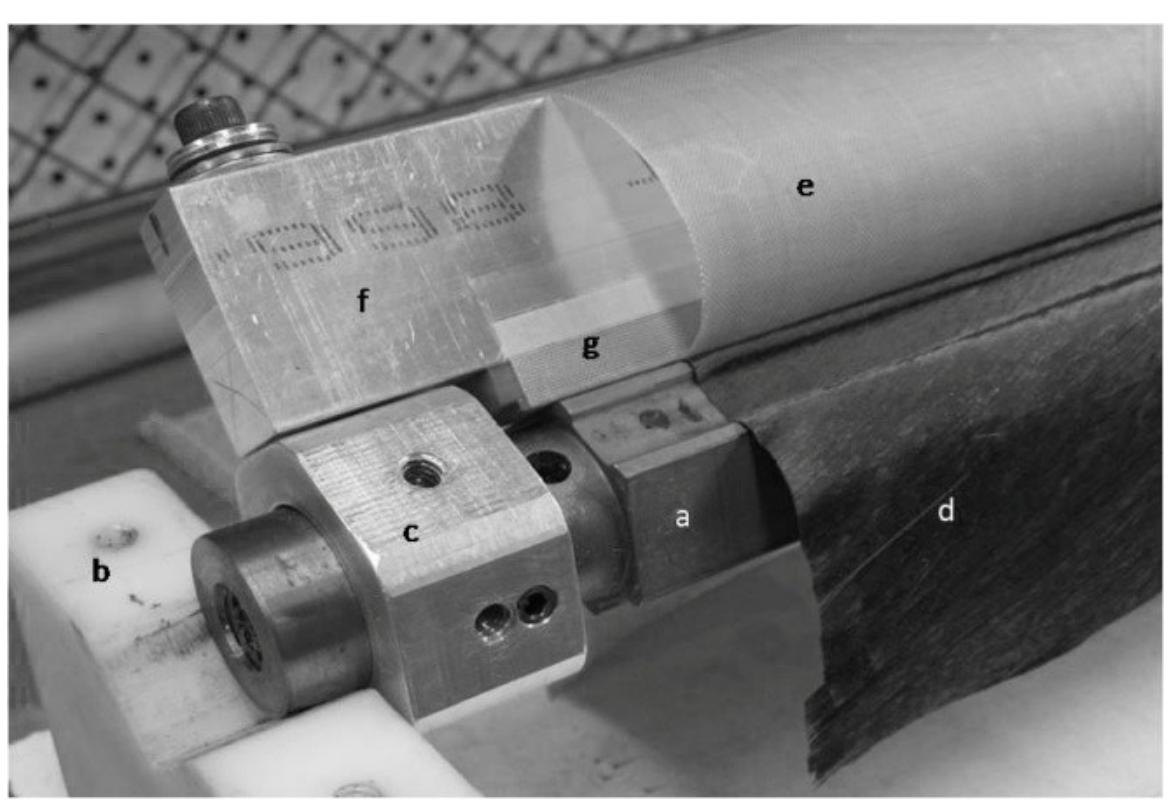

Figure 6. Material lay-up. a) mandrel, b) v-block, c) collar, d) pre-preg laminate, e) release fabric, f) clamping bar, g) silicone pad.

Wrapping the polyimide-Cu ground surface (described below) completes the lay-up and requires a different technique because the material comes in strips that cover only 120º circumferentially. A strip is put down to cover an upper surface so that it is oriented along the full length. To do this we first establish a precise centerline by lightly scribing a point at each end of the carbon fiber lay-up. A continuous centerline printed on the top surface of the polyimide-Cu strip is aligned with the scribe marks. A clamp bar is fixed to the top surface and the polyimide-Cu strip is carefully worked around the edges and fitted to the lower surfaces. Clamp bars are then fixed to the lower surfaces as shown in Figure 7.

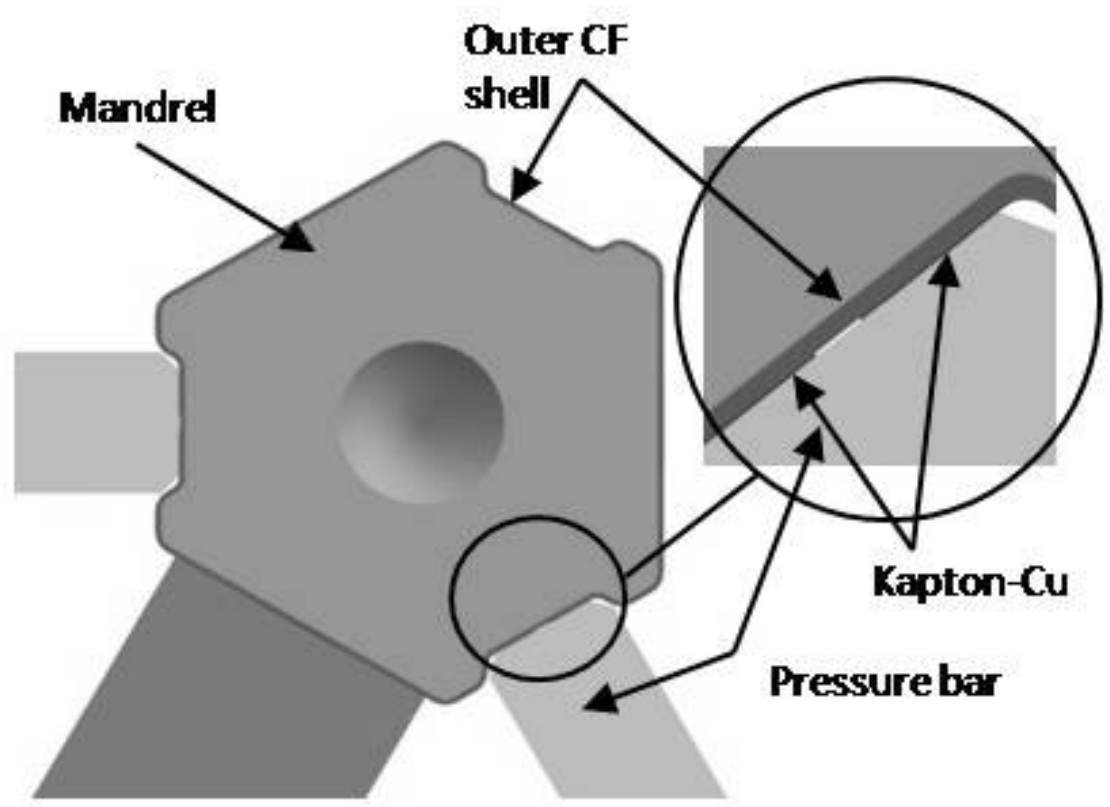

Figure 7. The use of three pressure bars holding the Kapton-Cu in place for co-bonding.

To position each of the adjacent strips, a clamping bar on one of the lower surfaces that contains a strip is removed, the next 120º piece is placed on an upper surface as previously described and its three bars fixed. The lay-up is complete when all three strips are attached and held by six clamping bars. The clamps are removed and silicone rubber pressure strips are placed in the three recessed regions. A protective silicone sheet 0.7 mm thick is wrapped tightly around the finished lay-up to prevent transfer of wrinkles from a vacuum bag, which goes on next. The structure is placed in an autoclave and cured. All lay-ups are cured for two hours in an autoclave at a pressure of 600 kPa at the manufacturer specified cure temperature of 135°C.

### 3.1.2 Inner Shell

The inner shell layup is simplified by first constructing a 0º/ 90º/ 0º ply. The 1.6m length and 12-facet geometry make the clamping system used for the outer shells impractical. In this case, the pre-laminate is laid out on a flat platen covered with a layer of release fabric. The mandrel is aligned with the leading edge of the ply and rolled forward while applying a moderate downward force. A 0.7 mm thick protective sheet of silicone is wrapped tightly around the lay-up and the structure is put in a vacuum bag and placed in an autoclave and cured. The autoclave pressure is maintained during the cool-down until the mandrel temperature has dropped below 70°C.

## 3.2 Cooling System

The PEEK cooling tubes are commercially extruded to our specifications. The material was not annealed after extruding, resulting in a less crystalline, near-amorphous molecular structure. This was done so that the tubes could more easily follow minor curves and transitions without buckling the walls. The cooling manifold components were CNC machined from natural, unfilled PEEK plate stock.

## 3.3 Support Structure Assembly

The various components of the support structure are joined using structural adhesives. The inner and outer shells are bonded to each other, and to the PEEK cooling tubes using thin film epoxy (3M® Aerospace, AF-163-2U, 0.14 mm thick) and cured at 100° C. The machined PEEK manifold parts are joined using a room temperature cure, liquid epoxy. In addition, electrically conductive silver epoxy (Loctite® 3888) is used to attach flexible grounding connections to the sensor shell polyimide/Cu mesh contacts (Figure 3).

### 3.3.1 Tooling

The required dimensional tolerances of the surfaces on which the sensors are attached are achieved by using precision tooling to adhesively bond the outer shells to the inner shell. The precision inner shell mandrel is used to provide reference surfaces for registering tooling components. Positioning bars attached to CNC machined collars mounted on the mandrel provide well defined and controlled envelopes that define the sensor shell surfaces while the adhesive cures.

### 3.3.2 Shell Assembly

The inner and outer shells and cooling tubes must be adhesively bonded continuously over their entire length with a well controlled glue bead. The thin film adhesive is supplied on a roll in a partially cured state, has moderate tack and is applied like tape to one of the mating parts. For the cooling tubes in the sensor region adhesive is applied to both sides of the Rohacell® foam spacers shown in Figure 3. The parts are then slid together and the cooling tubes lightly pressurized (10 kPa) to ensure good glue contact on all three sides during the cure.

The inner shell, sensor shells, and cooling tubes are bonded using the tooling to control the outer shell surfaces. In this operation the hybrid shells were temporarily installed to control and protect the cooling tubes and left unglued. Next the hybrid shells are bonded to the inner shell and cooling tubes. Tooling is not needed for this step because the hybrid mounting surface position tolerances are less stringent than the sensor mounting positions. The snug fit of the hybrid shells over the inner shell and cooling tubes is sufficient to spread the glue and position the parts.

Tubular jumpers are used to make a ground connection between the ground plane of the outer layer sensors, and the polyimide/copper ground circuit bonded to the carbon fiber support structure (Figure 3). There is one jumper for each outer layer sensor. The jumpers were fabricated from 10mm wide strips of polyimide with copper on one side. The strips are rolled and secured at the lap joint with adhesive to create a cylinder with the copper surface facing out. The cylinders are then pre-formed into the desired oval shape. Pre-forming the jumper eliminates the spring force that would otherwise act upon the sensor. The jumpers are first bonded to the ground plane on the support structure using electrically conductive silver epoxy. During sensor installation, silver epoxy is again used to connect the sensor ground plane to the jumper.

The hybrid circuits are mounted on aluminum nitride (ALN) rails in order to accommodate the cable stack that runs under them. These rails run the length of each hybrid circuit along both edges. The aluminum nitride is thermally conductive, and provides a heat path from the hybrid circuit to the support structure cooling system. One of the two rails of each hybrid is gold plated on three of its surfaces to provide a ground path between the hybrid circuit and the support structure. The gold plated ALN rails are bonded to the support structure with silver filled epoxy (TRA-DUCT 2902) and the bare ALN rails are bonded with thermally conductive ceramic filled epoxy (Masterbond EP30AN). The hybrid circuits are bonded to the rails with the same combination of silver filled and ceramic filled epoxies during the installation of the sensor-hybrid modules.

3.3.3 Cooling Manifold Assembly

The cooling manifold components include interlocking features that position the parts relative to each other. The parts were glued and assembled in situ to the inner shell and cooling tubes using a room temperature cure liquid epoxy (3M® DP-190). Extra care was taken to prevent excess glue from clogging manifold passages or cooling tubes. Once both manifolds were glued and assembled, the structure was loaded into a precision assembly jig for curing. When curing was completed the jig was used to establish the precision locations of the conical inserts described in Section 2.5 and to bond these in place.

## 4. Sensors

Four different sensor types, each 300 μm thick, are used in this detector. Narrow sensors have a nominal active width of 18.18 mm and have a 71 μm sense strip pitch. Wide ones have a nominal active width of 20.74 mm and have an 81 μm sense strip pitch. All sensor types have intermediate strips to improve spatial resolution. Each of these sensor types comes in two different lengths, 70 and 120 mm. The azimuthal coverage is 58.2° for the 18.18 mm sensor and 60.0° for the 20.74 mm sensor. Longitudinally, the 70 mm sensors are located closest to the center, two on each side of z = 0 and the 120 mm sensors are located at the outer ends of the active region. The 71 μm pitch inner sensors are radially closest to beam center (16.1 mm) and the 81 μm pitch outer sensors are at 17.6 mm. The sensor parameters and locations are summarized in Table 1. The sensor capacitance values listed in Table 1 are capacitances seen by the SVX4 chip inputs. The shortest analog cables were matched to the sensors with the highest capacitance in order to minimize the total input capacitance.

Table 1: L0 Module Parameters and Sensor Locations.

| z coverage (mm) | 0-70 | 70-140 | 140-260 | 260-380 |
|---|---|---|---|---|
| Sensor length (mm) | 70 | 70 | 120 | 120 |
| Sensor capacitance/strip (pF) | 8.4 | 8.4 | 14.4 | 14.4 |
| Analog cable length (mm) | 348 | 322 | 246 | 170 |
| Cable capacitance/strip (pF) | 11.1 | 10.2 | 7.8 | 5.4 |

| | Inner Position | Outer Position |
|---|---|---|
| Radius (mm) | 16.1 | 17.6 |
| Max angle (rad) | 0.51 | 0.53 |
| Strip pitch (μm) | 71 | 81 |
| Sensor Width (mm) | 18.18 | 20.74 |

Thirty sensors of each type (see Table 1) were purchased from Hamamatsu Photonics. All the sensors were tested at Fermilab using IV (current-voltage) and CV (capacitance-voltage) scans. Only three failed the leakage current specification near 800 volts.

Four sensors, one of each type, were sent to the State University of New York, Stony Brook for detailed strip testing. Sensors of all four types have a total of 511 strips: 256 readout strips separated by 255 intermediate strips. The chuck of an Alessi Rel 6100 probe station was used to apply positive high voltage to the backside of the sensor. A bias probe, mounted on the chuck, applied the ground connection. As many as five additional fixed probes contacted the AC and DC pads on neighboring strips and the DC pad on the intermediate strip between them. These probes all touched the appropriate pads on the sensor at one position of the chuck. A Hewlett Packard HP3904A Matrix Switch was used to avoid touching any pad twice. The chuck and sensor could step across a distance of several centimeters to an accuracy of about one micron with no significant variation of the probe marks on the sensor pads. Measurements were made of the following quantities for each strip:

$I_{diel}$ the dielectric current (nA) through $C_{ac}$
$C_{ac}$ the strip coupling capacitance (pF)
$I_{leak}$/strip the leakage current/strip (nA)
$R_{poly}$ the polysilicon resistance at one end of each strip (MΩ)
$Cond_{int}$ the interstrip conductance between two adjacent strips (pA/V)

Scans were made in pairs of strips.

Figures 8 and 9 show example scans of $I_{leak}$/strip and the interstrip conductance for each type of sensor. The quantization apparent in the measurements of the small currents is a measure of the systematic uncertainty in these measurements. Only one strip failed specifications (a pinhole due to a scratch) out of 1024 strips.

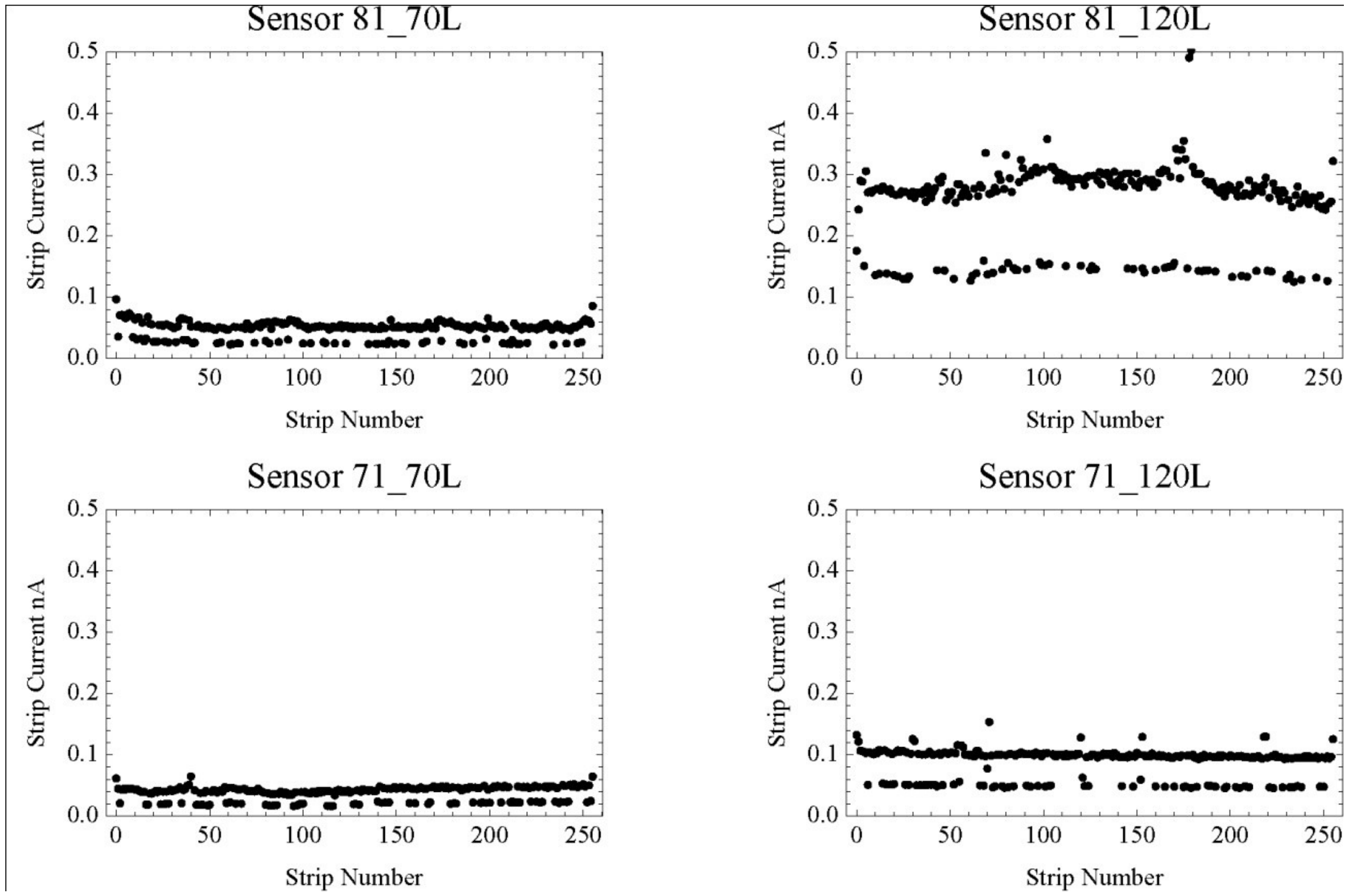

Figure 8: Strip scan results for the leakage current per strip ($I_{leak}$/strip) for each of the four sensors. The leakage current of one strip (41 in the 71_70L sensor) is off scale at 6.4 nA, but still within specifications.

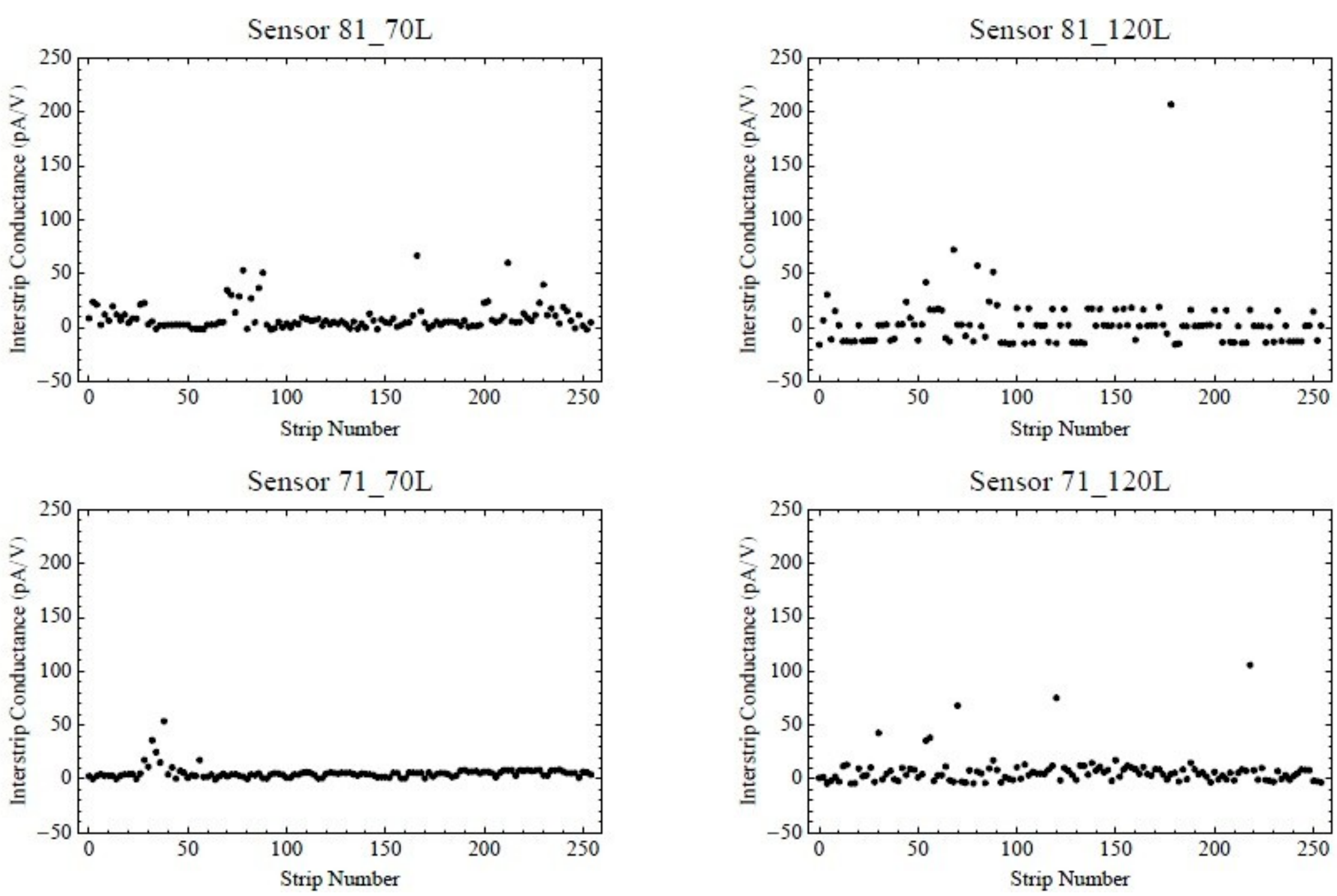

Figure 9. The interstrip conductance between adjacent strips.

All four test sensors were sent to Kansas State University (KSU) for irradiation in a 10 MeV proton beam at the KSU cyclotron. Two sensors received an exposure of protons corresponding to about 1 $fb^{-1}$ at the Tevatron and two sensors an exposure corresponding to about 4 $fb^{-1}$. Table 2 gives results for the parameters measured before and after irradiation. All Table 2 measurements and specifications are at room temperature. As expected the average leakage currents per strip for the irradiated sensors showed a substantial increase. The interstrip conductance increased substantially but was still less than the 500 pA/V limit. Scans of all strips of sensors showed that all quantities were reasonably uniform over all sensors, and thus well represented by the average per strip.

A definitive test of the interstrip resistance at the higher total exposures was not possible because a probe station at -10°C was not available. At these larger equivalent exposures the room temperature leakage currents were so high that the sensors could not be fully depleted in a CV scan due to the 50 kΩ resistors in series with the Source Measure Unit (SMU). These resistors were required in a CV scan to avoid interaction between the regulation of the SMU and the LCR meter. Consequently the measurements used the non irradiated Full Depletion Voltage (FDV) for these two sensors, since that was sufficient to fully deplete the sensors.

Table 2. Results of irradiation tests of sample L0 sensors in a 10 MeV proton beam. Values labeled 'ND' could not be determined due to high leakage currents. The 70 and 120 in the column headings identify the lengths of the sensors in mm.

| Sensor | P71_70L | | P81_ 70L | | P71_120L | | P81_120L | | Design Specs |
|---|---|---|---|---|---|---|---|---|---|
| Irradiation ($fb^{-1}$) | 0 | 0.8 | 0 | 3.2 | 0 | 1 | 0 | 4 | |
| FDV (Volts) | 190 | 150 | 190 | ND | 190 | 150 | 190 | ND | |
| $Cond_{int}$ (pA/V) | 5.5 | 275 | 9 | $2\times10^3$ | 7.0 | 390 | 3.3 | $1\times10^6$ | <500 pA/V |
| $R_{poly}$ (MΩ) | 0.82 | 0.71 | 0.92 | 0.86 | 0.94 | 0.81 | 0.92 | 0.72 | 0.8±0.3 MΩ |
| $C_{ac}$ (pF)/strip | 85 | 81 | 97 | 86 | 136 | 148 | 156 | 175 | >10 pF/cm |
| $I_{diel}$ (nA) | 0.05 | 0.003 | 0.15 | 0.01 | 0.12 | 0.13 | 0.33 | 0.003 | <10 nA |
| $I_{leak}$/strip (nA) | 0.06 | 220 | 0.05 | 1500 | 0.09 | 361 | 0.25 | 7000 | <10 nA at $V_{FDV}$ |

For the sensor exposed to an equivalent exposure of 3.2 $fb^{-1}$, scans showed that all quantities were again reasonably uniform over the sensor. However the sensor irradiated at 4 $fb^{-1}$ equivalent showed less uniformity. Because the FDV of these devices could not be determined, a complete study could not be performed.

# 5. Electronics Design

## 5.1 Design Requirements

L0 is an upgrade to an existing detector so it had compatibility as well as performance requirements. The new chip, referred to as SVX4 [2, 3] had to be compatible with both the SVX2 [5] and SVX3 [6] so that both CDF and D0 could use the chip with existing readout systems.

The requirements of the two experiments are similar enough that it was easy to agree on the number of channels per chip, input pad pitch, pipeline length and number of analog buffers. The difficulty was in the control lines. The SVX3 uses non multiplexed, differentially driven lines while the SVX2 uses multiplexed, single ended lines. There was not enough area for two sets of wire bond pads so we implemented a 'D0' bit that changed the function of some of the pads. This required building a translation board (called the 'adapter card') that converted the single ended D0 signals to differential signals for the SVX4 chip. The D0 mode bit also restricted the operation of the SVX4 to single mode operation. That is, only one mode could operate at a time so it was not possible to acquire data at the same time that previous data was either being digitized or read out. This operation is identical to the SVX2.

We obtained the electronics channels for Layer 0 by reusing the channels from the two H disks that were removed in order to install Layer 0. Although the SVX4 readout was quite similar to the SVX2 there were a few changes that had to be made to the existing electronics. As mentioned above, we had to build a new adapter card to accommodate the differential signals from SVX4. We also used this card to provide an isolated ground (described below) for the electronics mounted on the detector. The firmware in the readout sequencer had to be modified to account for slight differences in the SVX4 control logic and we added a small, passive card near the end of Layer 0 to serve as an interconnect point between the polyimide cables coming from the detector and the twisted pair cables going to the adapter card. The rest of the readout system was unchanged.

The small cross section and long length of the mechanical assembly had a large impact on both the signal to noise ratio and the detector's susceptibility to external noise. The small aperture meant that the readout chips could not be mounted near the sensors so they were connected to the sensors by polyimide flex cables which increased the input capacitance. The cables were matched to the sensors so that the overall capacitance was minimized but the best signal to noise ratio was only about 15 to 1 and it will decrease with increasing radiation damage.

The long length of the detector made it very difficult to provide a dielectric break at the center of the detector. Carbon fiber is electrically very conductive [7] so this lack of a dielectric break created a ground loop through the detector support structure. These features forced us to develop a very low noise design.

## 5.2 Noise Reduction Methods

We used two principal methods to reduce the detector sensitivity to common mode noise. The first was to isolate the detector ground from the rest of the world so that the potential ground loop through the detector was broken. Since we had to make a new card to provide translation between differential and single ended signals for the SVX4, we used the card to provide an isolated ground for the detector. Details of the card design are presented in the adapter card section.

Second, we tied all the detector elements together with low impedance connections so that the chip and the sensor saw the same reference voltage. This means that the front end amplifier reference (it's 'ground') is at the same potential as the sensor and the part of the detector that the input cables pass over. Since the amplifier measures the difference between its ground and the input signal, the noise signal is minimized.

The latter feature required an integrated detector structure that minimized the voltage difference between the SVX4, the hybrid, the detector mechanical structure and the sensors. One of the key elements to achieving this was making a good connection to the carbon fiber mechanical and cooling structure. This was achieved by using a thin copper clad polyimide sheet co-cured with the carbon fiber. The copper was etched to form a mesh ground plane with 234 µm lines spaced 1450 µm apart. The copper faced the carbon fiber and made a good electrical contact. Printed circuit technology was used to etch vias and gold plate pads on the reverse side at points where external connections had to be made.

One common problem that silicon detectors have is to make a low impedance return path from the ground of the preamplifier (in the SVX4) to the bias voltage side of the silicon sensor. This is especially true when a cable separates the amplifier from the sensor. The cable used in L0 has a DC resistance for the bias and bias return line of about 10 ohms each. These traces are side by side with no ground plane underneath so the AC impedance at operating frequencies is even higher. Such large impedance effectively isolates the sensor from the hybrid so that any stray fields coupling to either one will result in a voltage difference that will be seen as noise in the SVX4. This impedance was reduced by using the polyimide clad carbon fiber shell as the return path instead of the cable. The DC resistance is much less than 1 ohm and its large cylindrical shape minimizes the inductance.

The main issue then became the connection of the hybrid and the sensor to the mechanical structure. The hybrid was made with 3 vias through the BeO that connected the hybrid ground to a gold strip on the bottom surface. The carbon fiber mechanical structure has a set of gold plated aluminum nitride rails attached to the copper clad polyimide that matched the strip on the hybrid. The two gold strips were bonded together with conductive epoxy. The aluminum nitride was connected to the copper mesh on the other side of the polyimide by standard printed circuit vias.

The sensor connections were done in a similar manner. In this case a pitch adapter (described in Section 5.3) was used to provide a ground connection to the carbon fiber structure and to carry the bypass capacitor for the bias voltage. The ground connections to the wide sensors were made by the tubular jumpers described in the Section 3.3.2

Many connections in this detector were made with conductive epoxy. Because of concern about a failure of the conductive epoxy it was used it only to join parts that were already in direct contact. A very thin layer of conductive epoxy was used so that there was a high probability that there would be direct mechanical contact in addition to the epoxy contact.

### 5.3 Pitch Adapter

A small ceramic pitch adapter, mounted on the sensor surface, was used to adapt the pitch of the analog cables to the 71 and 81 μm pitches of the narrow and wide sensors (Figure 10). This allowed a single style of cable to service both detector types. The adapters also carry HV bias-to-ground decoupling capacitors and provide lands for the flex circuit jumpers that wrap around the detectors to provide +HV to the detector back side and carry the support structure ground to the detector surface. The adapters are 250 μm thick $Al_2O_3$ ceramic with gold over a titanium-tungsten base metallization.

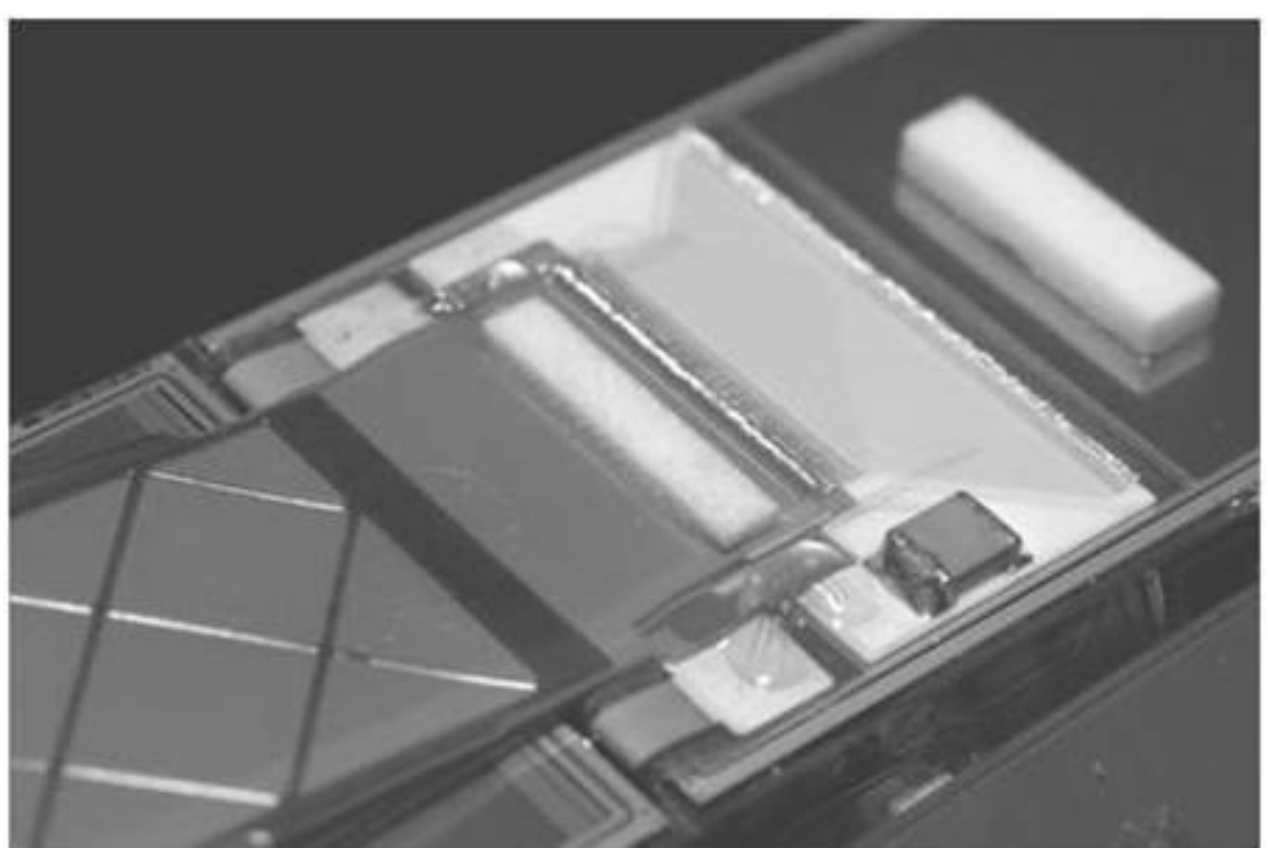

Figure 10. Pitch adapter showing the connection to the analog cable, the bypass capacitor and the tabs that connect the bias voltage to the back side of the sensor and detector ground to the capacitor.

Cables were glued to the pitch adapters and the assemblies were then glued to the sensor surface. Wire bonds were made from the detector strips to the pitch adapter and then from the pitch adapter to the analog cable. Foam bumpers were placed near the wire bonds to protect them from cables passing over the module.

### 5.4 Analog Cables

The L0 design employed pairs of analog cables, each with a width of about 13 mm. These pairs were made in four different lengths varying from 170 mm to 348 mm. Each cable of a pair (called 'top' and 'bottom') has a trace pitch of 91 µm and the top cable is displaced with respect to the bottom cable by half a pitch. The top and bottom cables are separated by a 90 µm thick polyimide mesh spacer that has 90% open area. They have a length difference of 2.15 mm to allow access to the bonding pads of the bottom cable in the assembly. The two cables were adhesively bonded together at each end such that the resulting assembly has an effective pitch of about 45 µm matching the pitch of the traces on the pitch adapter at the sensor end. Copper reinforced holes at four locations in the cable aided in the positioning and alignment of the cable pairs.

There are four stacks of these cables per detector facet. Each pair of cables is also separated from the others by a 90 µm thick polyimide mesh spacer. This spacer reduces cross talk from the other cables and adjacent hybrids. The bottom layer is separated from the carbon fiber support structure by four of these polyimide spacers.

The analog cables were produced[7] from 50 µm thick polyimide material with adhesiveless copper cladding. Each cable of the pair has 128 gold-plated copper traces that are 15 µm wide and have a resistance of about 1 ohm/cm. There is also one 100 µm wide HV (bottom cable) and one ground (top cable) trace with a resistance of 0.3 ohm/cm. A 25 µm thick solder mask coating is applied on top of the HV trace to hold up to 1000 VDC. The gold layer on the 8 µm thick copper traces has a typical thickness of about 1 µm. The standard Nickel plating that is normally applied between the copper and gold metal layers was not used because it made the traces too brittle. The cable assembly was modeled in a 2D field simulation program to optimize its design with respect to total and mutual capacitances. The total trace capacitance of the cables was measured to be 0.32 pF/cm, which is within 10 % of the simulated value. Two hundred individual analog cables were delivered with no defects.

## 6. SVX4 Hybrid Design

The circuit board carrying the SVX4 chips is a multilayer, thick-film ceramic hybrid. Six layers of metal alternating with five layers of dielectric are screen-printed onto a Beryllium Oxide substrate. BeO was chosen as a substrate mainly because of its low density and high thermal conductivity.

Several techniques were employed to optimize performance. Clock and data lines are Low Voltage Differential Signals (LVDS). To reduce the total mass, one layer is used for both analog and digital power. This plane doubles as a virtual ground and it has no interruptions for better signal referencing. The main ground and power layers use a mesh

[7] Dyconex AG, Switzerland

(0.04 in. wide traces spaced 0.01 in. on center) except where they pass directly under signal traces. The ground plane is a mesh everywhere except under the signal traces where it is solid metal.

The bottom surface of the SVX4 chip is and analog ground and it is important to minimize any current flowing in this ground. This was accomplished by making a small ground plane on the hybrid that is the same size as the chip. This plane is connected with three vias on one end to the main ground plane. Three vias are used in order to minimize the inductance. This configuration substantially reduces ground currents in the local ground plane resulting in reduced analog noise. In addition there are no digital traces in the region under the chips. The analog supply for each chip is bypassed to this small ground plane with a 0.01 μF capacitor. The digital supply for each chip is also bypassed to the main ground plane with a 0.01 μF capacitor. The back side of the substrate has a metal strip along one edge that is connected by three vias drilled through the substrate to the main ground layer. The strip is connected to the copper mesh on the support structure with silver epoxy.

### 6.1 Adapter Card

This card provides both the conversion from single ended to differential output as well as isolation of detector grounds. The isolation barrier needed to pass eight bi-directional data signals with transmission rates up to 26 MHz, a 53 MHz clock and five low speed control lines. We investigated several methods of ground isolation including optical couplers, magnetic couplers and single ended to differential translators. The optical couplers were too slow for the 53 MHz clock that we use. The magnetic couplers fail in the solenoid field. The single ended to differential translators were the only ones that met our requirements. Texas Instruments, Inc. SN65LVDM1677 was used to transmit the bi-directional data, data latch and control signals and a SN65LVDT100D was used to retransmit the system clock.

The board layout was optimized to reduce capacitive coupling of the electronics ground and the detector ground. This was accomplished by locating the ground planes on the same layer with as much spacing as component placement would allow. Additionally, power trace runs over the opposite ground reference were minimized. A 120 ohm resistor was connected between the two grounds so that the Low Voltage Differential Signals (LVDS) would have a common DC reference between the two grounds. A bare board has impedance between the two isolated grounds of 278 ohms at 7 MHz with little frequency dependence. A populated adapter card without cabling has an impedance of 73 ohms which was reduced to 33 ohms when all the cables were attached. There are 6 boards per end (the grounds are all in parallel) so the impedance between the grounds was a little less than six ohms. Note that this is the worst case result because the grounds on the non detector side are only connected together at the end of a 3 meter cable.

## 7. Installation

For each half of the Run IIa silicon tracker, existing coordinate measuring machine (CMM) data linked Run IIa sensor locations with openings in the support structure membranes at z =0 and at the outer ends of the each half. Data from a survey conducted in the fall of 2004 confirmed this earlier information, measured the aperture available at overlapping membrane openings at z = 0, established the position of that aperture relative to the openings at outer ends of the support structure, and showed that other intermediate membrane openings did not additionally constrain the aperture for L0 installation.

An installation concept was developed that involved inserting a long tool through the Run IIa SMT from one end, connecting this to L0 and then pulling L0 back through the SMT to its final position. A detailed written L0 installation procedure was developed and tested with the aid of a full scale mockup of the existing silicon detector support structure, a prototype L0 support structure with dummy sensors, and the proposed installation tooling. Multiple trial installations led to significant improvements in the procedure and tooling and ensured that all personnel were thoroughly familiar with the procedures to successfully install L0.

After the detector was opened, the first step was to establish an optical center line between the north and south ends of the SMT based upon the openings in the outermost membranes of the Run IIa silicon structures. Because the space for survey equipment was limited and a short focal distance was necessary, a Brunson model 76 transit square was selected and specially calibrated so that it could also be used as a level. A combination of precision bubble levels and transit stages equipped with digital micrometers allowed the Brunson to be positioned in a well-controlled and known way. The Run IIa beam pipe was removed using the installation tooling. Cantilevered support was provided to the beam pipe during most of the removal process. Gravitational deflection and transverse motion were monitored by the Brunson using cross-hairs inserted into the beam pipe flanges and appropriate adjustments were made to the support tooling as needed. The beam pipe was extracted into the larger diameter beam pipe of an end calorimeter and then removed from the outer end of the calorimeter later.
Mount rings for L0 were attached with epoxy to each end membrane of the SMT. Transverse locations of the mount rings were centered on the openings in the SMT end membranes. Azimuthal orientation was set relative to the gravitational direction using precision electronic levels (Advanced Orientation Systems, Inc, EZ-Tilt-3000, nominal repeatability < 3 $x10^{-4}$ arc-deg).

L0 was inserted via an end calorimeter beam pipe. Temporary spacers were attached to L0 so that it could slide through the beam pipe without damage to the sensors. Rails for the L0 installation tooling were installed and aligned at each end of the central calorimeter. Both tooling alignment and deflection were critical and required constant monitoring, with the Brunson, and adjustments during L0 installation When L0 began entering into the Run IIa silicon tracker, no Brunson observations were possible. It was necessary to rely on knowledge of the tool and L0 transverse locations as they were moved longitudinally into place. This was established by calculations and measurements made during the installation trials. To minimize tool deflections, a special "long" tool was fabricated from high modulus carbon fiber by the University of Washington. Space constraints required that the long tool be fabricated in two sections joined with a precise, reproducible connection. Deflection of the tool under gravity with cantilevered support and simulated L0 loads were established in advance on a CMM.

The long tool was inserted though the SMT aperture and its transverse motion was monitored with the Brunson during insertion. The tool was cantilevered from two stages (Figure 11) carried by the installation rail system. Longitudinal space constraints required that the tool be inserted in several steps. When one stage reached the end of the rail, a third stage was attached to the tool's outboard end and the innermost stage was then removed. The process was repeated until the tool extended fully through the SMT. Dial indicators monitored the tool transverse motion at each stage of this operation and allowed for predetermined compensations for tool gravitational deflection.
L0 was secured to similar stages at the other end of the detector in preparation for installation. The initial elevations and transverse positions of those stages were set using the Brunson. After the long tool and L0 were in position, the two were coupled together. Final installation consisted of withdrawing the long tool and thus pulling L0 into the SMT aperture.. Adjustments in stage elevations and multiple transfers of tool and L0 grip points were required during the process and monitored by dial indicators. Once L0 was in its final position, G-10 rings, studs, washers, and nuts were used to connect precision cone and ball mounts on L0 to the precision cone and ball mounts of the mounting rings previously glued to the SMT end membranes. This completed the mechanical installation of the main L0 assembly.

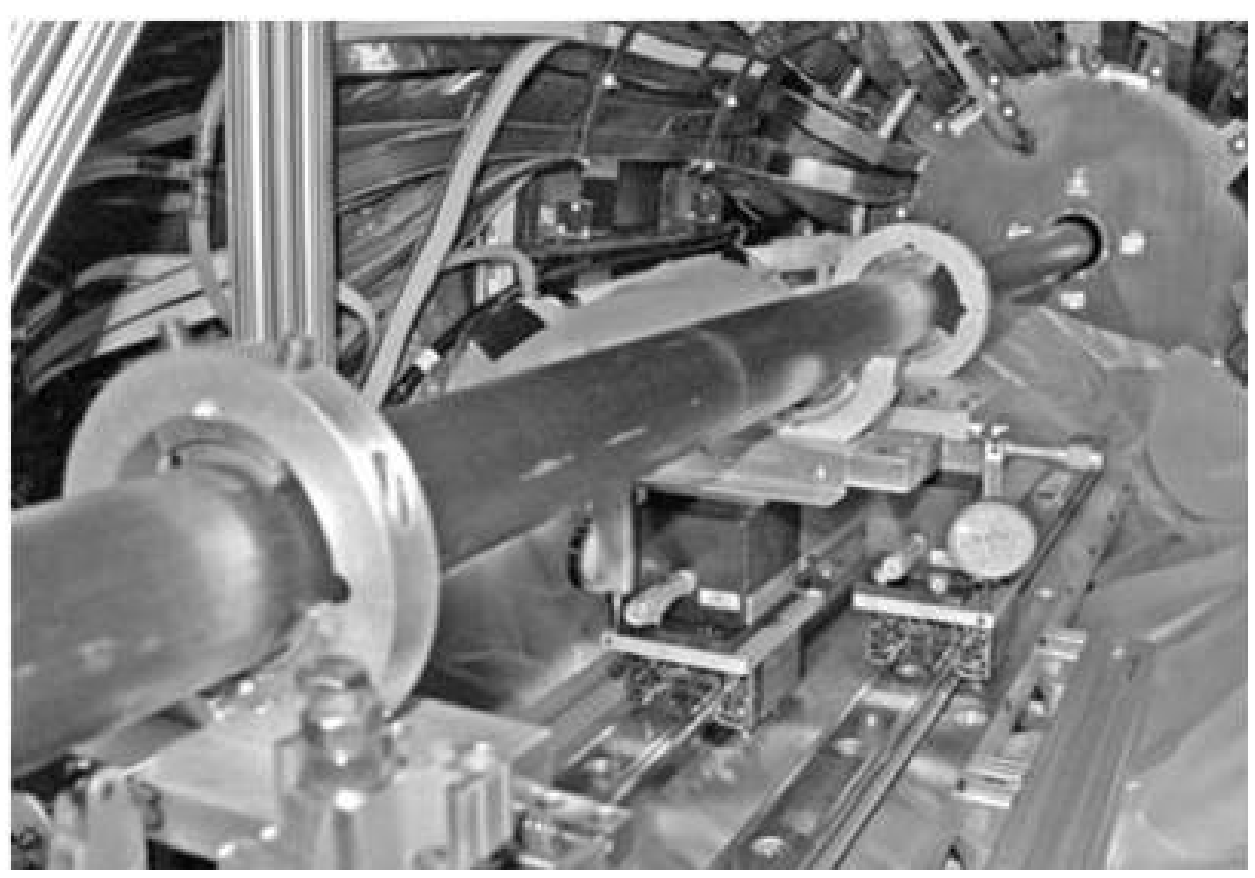

Figure 11. Insertion tool at the South end of the SMT.

Subsequently, the new beryllium beam pipe was inserted through the L0 support structure by hand and supported from the L0 structure by insulating kapton bands. L0 is designed to have an isolated ground so it is important that the detector is centered around the beam pipe. This was checked by measuring the capacitance between the beam pipe and the detector. The value was within 10% of what we calculated for two concentric cylinders at the same relative positions and agreed well with bench measurements. This confirmed that the beam pipe and L0 grounds were isolated and that the beam pipe was sufficiently centered within L0.

Mounts for the junction cards were attached to the SMT end membranes with epoxy, junction cards were mounted, and cabling was completed. Coolant connections were made and flow of room temperature coolant was initiated so that multiple checks of L0 readout and cable assignments could be made. Beam pipe extensions and associated supports were installed, connections to the end calorimeter beam pipes were completed, thermal insulation and membranes to provide a dry gas enclosure were installed and flow of dry purge gas was initiated. The full L0 installation was successfully completed on schedule with no undesired events.

# 8. Performance

## 8.1 Initial Operation

The L0 commissioning began immediately after installation and completion of the electrical connections. Stable, error-free readout was achieved for all 96 SVX4 chips in L0. The total number of bad channels was less than 20 out of 12288, all due to defective wire bonds made during module production.

## 8.2 Performance

The new detector has significantly improved the physics performance of the D0 detector. Since it was designed to improve the vertex resolution, it is not surprising that the biggest gain is in the measurement of the impact parameter. Figure 12 shows the impact parameter resolution with and without the Layer 0 detector.

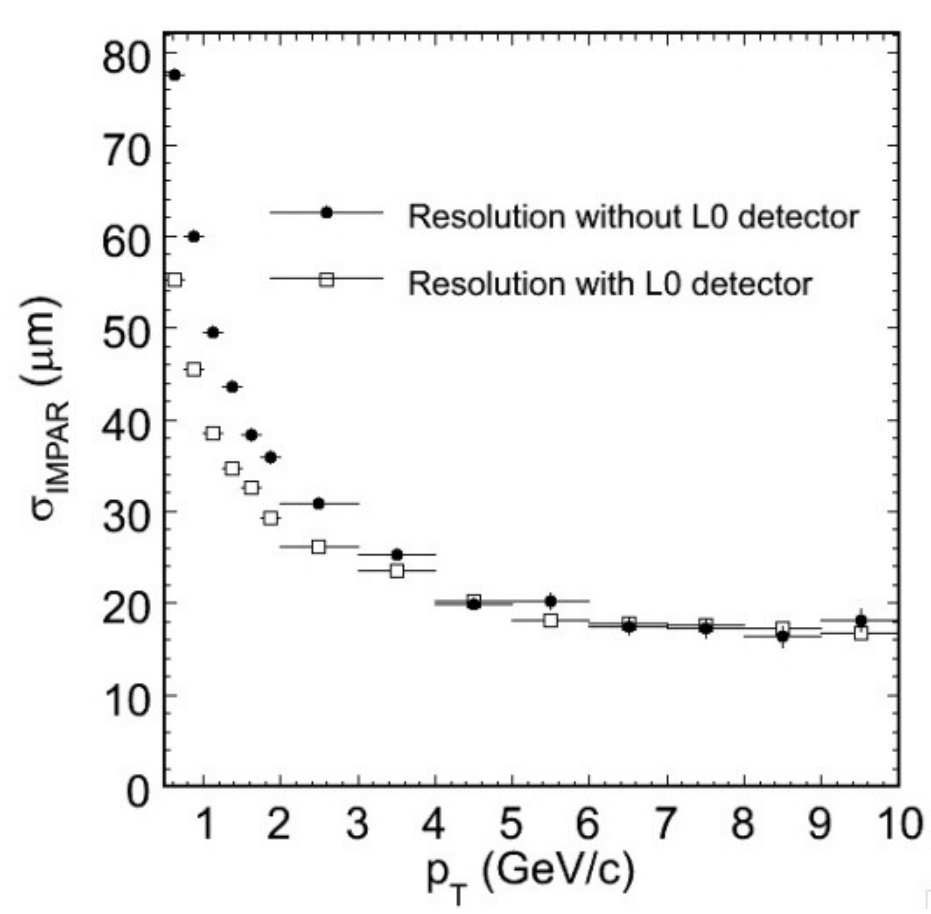

Figure 12. The impact parameter resolution with and without the Layer 0 detector.

The distribution accounts for the L0 intrinsic resolution (strip pitch), position of first hit, multiple scattering and vertex position uncertainty. For $P_T$ values less than about 2 GeV/c the improvement in resolution is greater than 20%

### 8.3 Signal and Noise

An outstanding noise performance was observed with both L0 and SMT included in the detector read out. No evidence of cross talk was observed between L0 and other parts of silicon detector.

Figure 13 shows typical pedestal, total noise per channel and differential noise for one of the L0 modules in normal running mode. The average differential noise is about 1.8 ADC counts and the average total noise per channel is about 1.7 ADC counts. From Appendix A, the single-channel noise is expected to be 1.4 ADC counts, so we conclude that there are about 0.3 ADC counts of common mode noise.

The amplitude for a minimum ionizing particle is 25 ADC counts. Therefore an average signal-to-noise ratio of 15:1 is obtained for hits in L0. The pedestals are fairly uniform across all the chips. L0 is read out in the zero suppressed mode where only channels above a threshold and its two nearest neighbors are read out. Noise and pedestals are very stable in time.

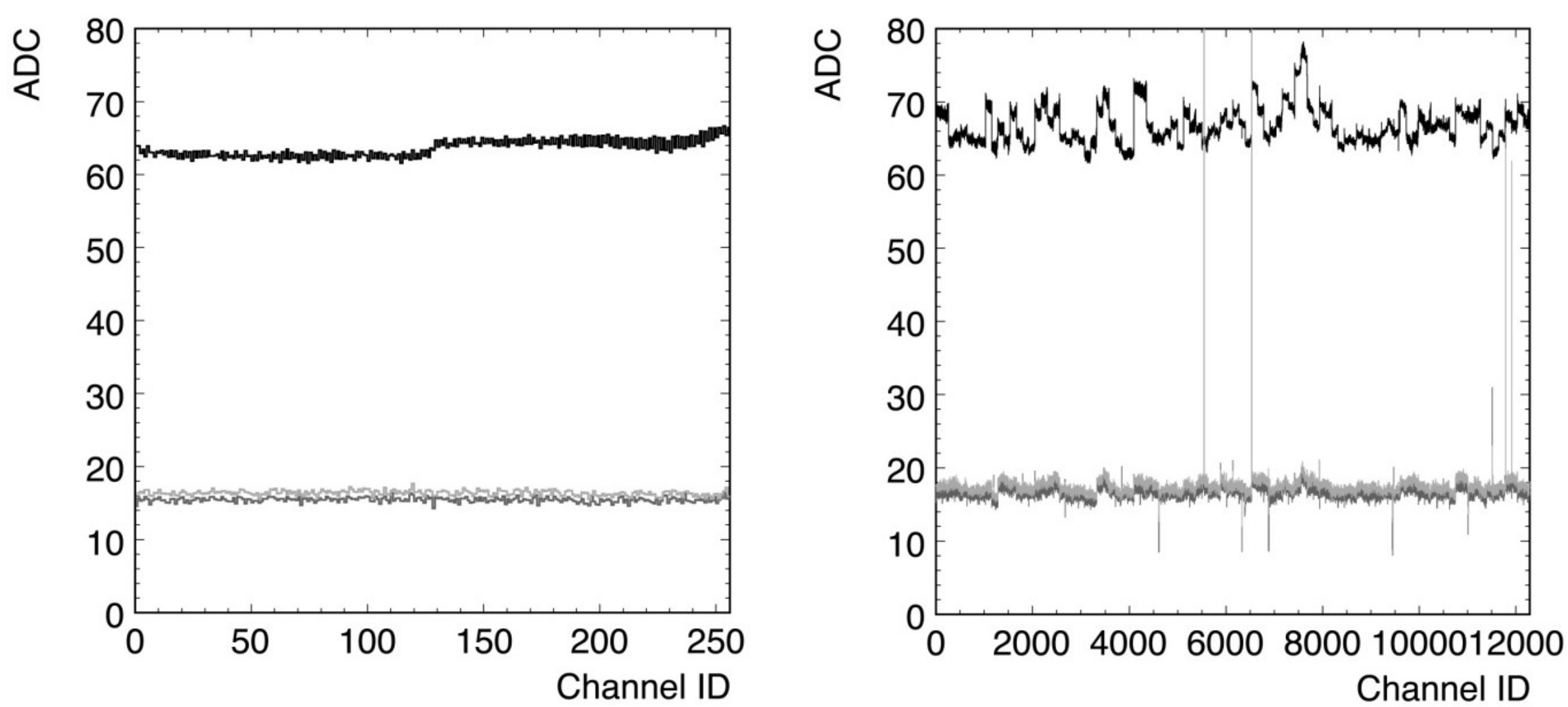

Figure 13: The left plot shows typical pedestal (black histogram), total noise per channel×10 (dark gray histogram) and differential noise×10 (light gray histogram) as a function of channel number for one Layer 0 module after installation inside the SMT. The right plot shows the same distribution for all 96 chips.

**Summary**

The major contribution of this paper is the development of new methods of using carbon fiber for both the low-Z mechanical structure and the ground connections in the L0 silicon detector. The paper presents a detailed description of the techniques to co-bond a thin copper clad polyimide layer to high modulus carbon fiber. This polyimide layer forms a good electrical connection to the carbon fiber so that the entire structure acts like a low impedance ground connection between the various electrical components. Such a low-impedance ground connection is particularly crucial between the sensors and the readout chips. This ground connection coupled with isolating the local SVX4 grounds from the rest of the world reduces the common mode noise to 0.3 ADC counts (1% of a MIP).

The detector was installed in the summer of 2006 and has since performed very well and resulted in an improved impact parameter resolution. The only failure that has occurred is that the high voltage line to one sensor has opened up resulting in the loss of this sensor.

## Appendix A: Noise Calculations

We define the noise on the output of channel *n* as the standard deviation of the digital output, *a*, in ADC counts for that channel:

$$\sigma_{a,n}^2 = \langle a_n^2 \rangle - \langle a_n \rangle^2 \tag{1}$$

We also define "differential" noise on a given channel [8] as

$$\left(\sigma_a^{dff}\right)^2 = \frac{\left[\left\langle \left(a_{n+1} - a_n\right)^2 \right\rangle - \left\langle a_{n+1} - a_n \right\rangle^2\right]}{2} \tag{2}$$

which is estimated by computing the root mean square value of

$$\frac{a_{n+1} - a_n}{\sqrt{2}} \tag{3}$$

for a large number of events.

Differential noise would be a good estimator of common mode noise if there were no correlations between adjacent channels. However, adjacent channels are correlated both through the inter-strip capacitance and the capacitance between traces in the analog cable. These correlation effects need to be taken into account when evaluating Equation (2). Defining $Q_n$ as the total noise charge (random plus common mode) in a given channel, n, it follows that $\langle Q_n \rangle = \sigma_{a,n}$, where $\langle Q_n \rangle$ is the expectation value of $Q_n$.

Following [8] we get the total noise per channel charge, $Q_n$ , in channel n is

$$\begin{aligned} Q_n &= C_s(U_{n-1} - (2C_s + C_g + C_A)U_n + C_sU_{n+1} \\ &= C_s\left(U_{n-1} - (2+r)U_n + U_{n+1}\right) \end{aligned} \tag{4}$$

where $U$ is the noise amplitude, $C_s$ is the capacitance between adjacent channels including the analog cable, $C_g$ is the capacitance to ground, $C_A$ is the total amplifier capacitance and $r = (C_g + C_A)/C_s$. The ratio of the differential noise to the noise in a single channel is:

$$\frac{(Q_{n+1} - Q_n)}{2Q_n} = \frac{(U_{n+2} + (3+r)(U_n - U_{n+1}) - U_{n-1})}{2[U_{n-1} - (2+r)U_n + U_{n+1}]} \tag{5}$$

Assuming that all the noise amplitudes are identical and converting to expectation values gives:

$$\frac{\langle (Q_{n+1} - Q_n)/2 \rangle^2}{\langle Q_n \rangle^2} = \frac{10 + 6r + r^2}{6 + 4r + r^2} \tag{6}$$

$C_A$ is estimated [9] from the noise performance of the SVX4 (380 +41 electrons/pF) to be 9.3 pF. From sensor and analog cable measurements, $C_i$ = 11 pF and $C_g$ = 1 pF giving r = 0.93 and

$$\langle Q_n \rangle = 0.8\langle (Q_{n+1}) - Q_n \rangle/2 \tag{7}$$

Consequently, the expected total noise per channel is 80% of the differential noise that is calculated in (2). If the single channel noise is larger than this, there is a contribution from common mode noise.

**Acknowledgments**

We would like to thank the Department of Energy and the National Science Foundation for support during the course of this work and acknowledge the many contributions of the University of Washington Physics Department Machine Shop and the technical Staff at the Fermilab SIDET laboratory. We also thank Jim Fast for contributions during the conceptual development of L0 and Meghan Anzelc and Selcuk Cihangir for contributions during the installation. We are grateful to Jon Kotcher, the RunIIb upgrade manager, Vivian O'Dell, who replaced him, and George Ginther for their support and encouragement. We thank our D0 colleagues for providing the supporting infrastructure and for many interesting discussions.

**References**

[1]. The D0 RunIIa Silicon Microstrip Tracker, S.N. Ahmed et al., (to be submitted to Nucl. Inst. and Meth A).

[2]. B. Krieger et al., IEEE Trans. Nucl. Sci. 51 (2004) 1968.

[3]. M. Garcia-Sciveres et al., Nucl. Inst. and Meth. A 511 (2003) 171.

[4]. C. H. Daly et al., Design and Analysis for the Carbon Fiber Composite Structure for Layer 0 of the D0 Silicon Micro Tracker, D0 Note 5333. Available as Technical Memo FERMILAB-TM-2444-PPD.

[5]. T. Zimmerman et al., IEEE Trans. Nucl. Sci. 42 (1995) 803.

[6]. T. Zimmerman et al., Nucl. Inst. and Meth. A 409 (1998) 369.

[7]. W. Cooper et al., Nucl. Inst. and Meth. A 550 (2005) 127.

[8]. G. Lutz, Nucl. Instr. and Meth. A 309 (1991) 545.

[9]. T. Zimmerman, private communication.